\documentclass[CRPHYS,Unicode,manuscript]{cedram}
\usepackage{graphicx} 
\usepackage{bm}
\usepackage{physics}

\title{Towards analogue black hole merger}
\author{\firstname{Dmitry} \lastname{Solnyshkov}\IsCorresp}
\address{Institut Pascal, PHOTON-N2, Universit\'e Clermont Auvergne, CNRS, Clermont INP, F-63000 Clermont-Ferrand, France.}
\address{Institut Universitaire de France (IUF), F-75231 Paris, France}
\email[D. Solnyshkov]{dmitry.solnyshkov@uca.fr}

\author{\firstname{Ismaël} \lastname{Septembre}}
\addressSameAs{1}{Institut Pascal, PHOTON-N2, Universit\'e Clermont Auvergne, CNRS, Clermont INP, F-63000 Clermont-Ferrand, France.}

\author{\firstname{Guillaume} \lastname{Malpuech}}
\addressSameAs{1}{Institut Pascal, PHOTON-N2, Universit\'e Clermont Auvergne, CNRS, Clermont INP, F-63000 Clermont-Ferrand, France.}

\begin{abstract} 
  We study the effects of the wavevector-dependent losses on polariton condensates. We demonstrate that because of these losses, a single vortex becomes a center of a convergent flow, which allows describing it by an analogue Kerr black hole metric with a dynamically evolving origin. For a pair of vortices, we find an analogue of the 3rd Kepler's law and estimate the emission rate of the gravitational waves. We simulate an analogue of the inspiral phase of a black hole merger. Our work therefore suggests that polariton condensates with quantum vortices represent a setting with a fully self-consistent dynamical metric for broad analogue studies.
\end{abstract}
\begin{document}

\maketitle

\section{Introduction}
Analogue physics is based on the similarities between the mathematical models describing different systems, where the solutions found in one field can be used in the other fields. This cross-inspiration has already helped to discover the Higgs field and its excitation, the Higgs boson, by analogy with symmetry breaking in superconductors \cite{Anderson1963}. It has also helped to explain the extraordinary properties of graphene by Klein tunneling \cite{Katsnelson2006,Huard2007,Stander2009}, an effect predicted in high-energy physics  \cite{Klein1929}, but never observed in its original form.
There is currently an important activity in a particular branch of analogue physics: the analogue gravity \cite{Barcelo2005,Barcelo2018}. The topic of  analogue black hole simulations was started by the seminal work of Unruh \cite{Unruh1981}, that focused on an analogue of the Hawking emission \cite{Hawking1974}, originally expected to occur at the astrophysical black hole horizons, but then predicted by analogy and observed experimentally~\cite{Weinfurtner2011,Steinhauer2014} at subsonic-supersonic transitions in classical and quantum fluids.

The first studies on the analogue gravity were considering conservative fluids, both classical \cite{Unruh1981} and quantum \cite{Garay2000}. The presence of a steady fluid flow therefore required the presence of a drain (and a source). In 1D, the fluid is usually injected at one edge of the system and removed at the other edge \cite{rousseaux2008observation,Weinfurtner2011}. In 2D, the so-called "draining bathtub" configuration is used, with a local drain in the center \cite{Torres2017,sv2023exploring}. Quantum fluids, such as atomic BECs (Bose-Einstein condensates), were rather considered in 1D, without losses, in a dynamical configuration with a moving potential \cite{Lahav2010,Steinhauer2014}. The direct effect of the losses on the Hawking emission in analogue systems started to be considered more recently \cite{Robertson2015}.

Exciton-polariton (polariton) condensates \cite{kasprzak2006bose}, formed from strongly coupled excitons and photons in microcavities\cite{Microcavities} under sufficiently strong pumping, are a well-known example of a lossy interacting quantum fluid \cite{carusotto2013quantum} showing superfluidity \cite{amo2009superfluidity}, topological defects \cite{lagoudakis2008quantized}, and allowing to generate a wide variety of flows. Polariton lifetime usually ranges from 0.2 to several hundreds of picoseconds \cite{Levrat2010,nelsen2013dissipationless,gianfrate2020measurement}. This lifetime may seem shorter than in other systems (e.g. evaporation times of cold atoms or liquid helium), but polaritons also exhibit much faster dynamics, both in their propagation (thanks to their small effective mass, $m_{pol}\sim 5\times 10^{-5}m_0$ with $m_0$ being the free electron mass) and in their energy relaxation (thanks to efficient polariton-polariton and polariton-phonon interactions), the latter allowing, in some cases, to achieve thermal equilibrium in spite of the short lifetime \cite{kasprzak2008formation,Levrat2010}. In analogue gravity studies, these losses have mostly been considered as being wavevector-independent: they are usually described mathematically by terms like $-i\Gamma(x,y)\psi(x,y,t)/2$, where $\Gamma(x,y)$ is the polariton decay rate (with a possible spatial profile, allowing to describe the draining bathtub). If a polariton flow is superfluid at some point, these losses necessarily lead to the formation of a horizon \cite{Solnyshkov2011,Gerace2012}. This enabled the demonstration of the formation of a horizon in a 1D polariton wire experimentally \cite{HaiSon2015}. The possibilities to measure the dispersion of the excitations of the polariton condensates in order to extract the speed of sound, as well as to obtain the flow velocity from the phase of the condensate wavefunction, have inspired a lot of experimental activity \cite{boulier2020microcavity,jacquet2020polariton,jacquet2022analogue,claude2023spectrum}. 2D polariton superfluids may also enable the study of analogue gravity through the prism of Andreev-Hawking phenomena~\cite{zapata2011resonant,septembre2022angular}. Particularly important for the present work are the studies of vortex dynamics, which have been already carried out in polariton condensates \cite{Nardin2011,lagoudakis2011probing,manni2013spontaneous,gnusov2023quantum}. 

So far, whatever the platform, in most of the studies of analogue gravity the metric was static, determined by the conditions of the experiment and unable to react dynamically \cite{Barcelo2005}. For example, in the studies of superradiance \cite{Braidotti2020,Braidotti2022}, the energy of the black hole was decreasing due to the accumulation of negative-energy Bogoliubov excitations, but the metric itself was not modified by this accumulation because that would represent a higher-order correction. The examples of dynamical effects include an analogue black hole ringdown, that is, a decay of a weak perturbation of the horizon, that has been demonstrated recently \cite{Patrick2018}.
In particular, in the studies of backreaction \cite{Fischer2005,Balbinot2005,liberati2020back,Patrick2021}, all changes to the metric remain weak. In general, most of the settings allow simulating a "large" black hole and a weak excitation propagating in its curved spacetime, but not a strong perturbation or a modification of this spacetime. Analogue Penrose effect with quantum vortices \cite{solnyshkov2019quantum} has allowed to strongly modify the angular momentum of a black hole and therefore its metric, but only partially because the black hole's attraction remained the same. Meanwhile, the ultimate goal of achieving emergent analogue gravity \cite{barcelo2001analogue,Liberati2006,liberati2017analogue} obviously requires a configuration with a metric capable of responding dynamically and evolving with time.

The most well-known dynamical process already observed experimentally with real black holes is the so-called black hole merger \cite{LIGO2016}. It involves two black holes on a close orbit, spiraling in due to the emission of gravitational waves \cite{einstein1916naherungsweise,einstein1918gravitationswellen}. The inspiral process is followed by the merger of two horizons into a single one and then by a ringdown of this horizon to its final shape \cite{Price1994,blanchet2014gravitational}. The whole process requires a dynamical spacetime for its description:  the motion of a black hole means that the overall spacetime changes; the merger and the ringdown of the horizons imply a significant transformation of the shape of each of the two horizons, with released energy which starts to become comparable with that of a single black hole \cite{Pretorius2005}.

In this work, we demonstrate that the wavevector-dependent losses inherent to polariton condensates make a single quantum vortex in such condensate a center of a convergent flow, and thus make it an analogue black hole (more precisely, a Kerr black hole, because of the angular momentum of the quantum vortex).
Capable of motion, these quantum vortices thus represent a model system for studying the dynamics of a system of gravitating bodies (analogue black holes). We numerically consider a pair of same-sign vortices and describe their motion with an analogue "post-Newtonian" formalism. We derive modified analogues of the Newton's law of universal gravitation and of the 3rd Kepler's law, and find the solutions of the equations of motion for the inspiral stage. We compare these results with the well-known expressions for astrophysical black holes. The validity of our analytical solutions is confirmed by numerical simulations.

\section{Results and discussion}

\subsection{The model}

We will be dealing with polariton condensation under non-resonant pumping \cite{kasprzak2006bose,kasprzak2008formation}. As said in the introduction, the polariton losses due to the finite quality factor of the cavity and to the finite exciton lifetime have usually been described as being wavevector-independent.
This is, however, just an approximation. The polariton decay rate reads:

\begin{equation}
\Gamma_k=\abs{x_k}^2\Gamma_x+\left(1-\abs{x_k}^2\right)\Gamma_c
\label{pollifetime}
\end{equation}
where $\abs{x_k}^2$ is the exciton fraction of the polariton, $\Gamma_x$ the exciton lifetime and $\Gamma_c$ the cavity photon lifetime, which are both assumed to be constant, which again is an approximation \cite{richter2019voigt,su2021direct,krol2022annihilation}.

Near zero exciton-photon detuning, and for small $k$, the exciton fraction dependence on the wave vector can be approximated by:
\begin{equation}
    \abs{x_k}^2\approx 1/2+\frac{\hbar k^2}{2m_\mathrm{c}\Omega}
\end{equation} where $\hbar \Omega$ is the Rabi splitting and $m_{c}$ is the photon mass in the cavity. As discussed in~\cite{Solnyshkov2014}, this gives:
\begin{equation}
    \Gamma_k\approx \frac{\Gamma_x+\Gamma_c}{2}+\frac{\hbar k^2}{2m_{c}\Omega}\frac{\Gamma_x-\Gamma_c}{2}.
\end{equation}

Depending on the sign of $\Gamma_x-\Gamma_c$, a convergent flow can be realised using different methods. If $\Gamma_x-\Gamma_c>0$, the decay increases with $k$. In other words, the decay increases when getting closer to the core of the vortex, which provokes the convergent flow that can lead to the formation of an event horizon and to an attractive interaction between vortices. This condition can be realized in high-Q cavities or structures with fast excitonic decay. 

A more standard situation corresponds to $\Gamma_x-\Gamma_c<0$. In this case, a convergent flow can be realised by considering non-resonant pumping leading to the formation of a Bose-Einstein condensate.  For the condensate and for its excited states, the total net decay or growth rate $\eta_k$ includes the rates of scattering in $W_{in,k}$ and out $W_{out,k}$ of the modes:
\begin{equation}
\eta_k=W_{out,k}-W_{in,k}+\Gamma_k
\label{polgain}
\end{equation}
As analysed in \cite{Solnyshkov2014}, $W_{out,k}-W_{in,k}$ also scales as $k^2$ and the net scattering rate $\eta_k$ reads:
\begin{equation}    
    \eta_k\approx\eta_0+\Lambda\frac{\hbar^2 k^2}{2m_{pol}},
\end{equation}
where $m_{pol}$ is the polariton mass. 
The presence of a stationary condensate in the ground state means a zero net rate for the corresponding state: $\eta_0=0$. It also implies the absence of gain for higher-energy states, that is $\Lambda<0$, which is precisely equivalent to having $k^2$-losses.
Such dependence of the losses on the wave vector has been first introduced phenomenologically as the only way to reproduce experimental results \cite{Wertz2012,Anton2013}, providing a very good agreement with them.  Losses depending on the total particle energy were previously suggested to describe relaxation in superfluids \cite{Pitaevskii58} and Bose-Einstein condensates \cite{Choi1998,Wouters2010b}. However, the works on polaritons and polariton condensates have introduced a different shape of the loss term, depending explicitly on the kinetic energy only ($k^2$). Recent experiments have confirmed this $k^2$-dependence of the polariton decay \cite{fontaine2022kardar}.

We therefore study the behavior of the polariton condensate described by the  Gross-Pitaevskii equation with $k^2$-losses and saturated gain:
\begin{equation}
i\hbar \frac{{\partial \psi }}{{\partial t}} =  - \left( {1 - i\Lambda } \right)\frac{{{\hbar ^2}}}{{2m}}\Delta \psi  + U \psi  + \alpha {\left| \psi  \right|^2}\psi  + i\gamma {e^{ - {n_{tot}}/{n_0}}}\psi
\label{gpeloss}
\end{equation}
where $m$ is the polariton mass, $U$ is the confining potential (for example, a cylindrical mesa), $\alpha$ is the polariton-polariton interaction constant, $\Lambda$ is the constant characterizing the $k^2$-losses, and $\gamma$ is the prefactor of the gain term, saturated by the total density $n_{tot}$. We do not describe local reservoir saturation \cite{Wouters2007,Keeling2008,Wouters2010}, because we consider the case when the condensate dynamics is much faster than that of the reservoir.

Since the losses depend on the wave vector, and the largest wave vectors are present in the core region of the quantum vortex solution for a conservative condensate, it is natural to expect that the effect of the losses appearing in the modified equation~\eqref{gpeloss} will also be concentrated in the core region. The relation between the wave vector variation of the excitonic and photonic fractions and the spatial variation of the properties for a quantum vortex was already understood in Ref.~\cite{Voronova2012} (with consequences for the spatial density distribution) and used very recently in the description of the vortex kinematics via the Berry curvature in a recent work \cite{dominici2023coupled}, but the excitonic core in that work had a longer lifetime, and so the vortices were repelling each other.

In the following sections, we solve Eq.~\eqref{gpeloss} numerically to find the properties of a single vortex. We then look for an analytical description of the flows created by this vortex, taking the solution for a quantum vortex in the conservative case at large $r$ as a starting point and treating the loss term as a perturbation. We then study the behavior of a pair of vortices both numerically and analytically.

\subsection{Single vortex}
The metric of the weak excitations of vortex flow with non zero radial velocity $v_r$ is formally similar to the one of a Kerr Black hole \cite{Visser1998,Popov1972,fischer1999motion,solnyshkov2019quantum}. Indeed, a cylindrically-symmetric configuration with radial and azimuthal flows ($v_\phi=\pm\hbar/(mr)$), with an appropriate change of coordinates \cite{Berti2004}, allows writing the metric of the condensate as:
\begin{equation}
\label{Kcm}
g_{\mu\nu}=\frac{mn}{c_s}
\begin{pmatrix} 
-(c_s^{2}-{v_\mathrm{tot}}^{2}) & 0 & -2rv_\phi\\
0 & \bigg(1-\frac{v_r^2}{c_s^2}\bigg)^{-1} & 0 \\
-2rv_\phi & 0 & r^2\\
 \end{pmatrix}
 \end{equation}
with the local speed of sound $c_s=\sqrt{\alpha |\psi|^2/m}$, and $v_\mathrm{tot}$ the total velocity.

The event horizon is given by the condition $v_r=c_s$, whereas the static limit occurs when $v_{tot}=c_s$. In a conservative case, $v_r=0$, so there is a static limit, but no event horizon. We first perform a numerical solution of the modified Gross-Pitaevskii equation~\eqref{gpeloss} to find the stationary configuration with a single vortex and analyze its properties. We use the parameters typical for a GaAs cavity: $m=5\times 10^{-5}m_0$, $\alpha|\psi|^2\approx 1$~meV (for a stationary homogeneous solution) giving the healing length of the order of 1~$\mu$m, and the $k^2$-decay coefficient $\Lambda=0.1$.

\begin{figure}
    \centering
    \includegraphics[width=0.9\linewidth]{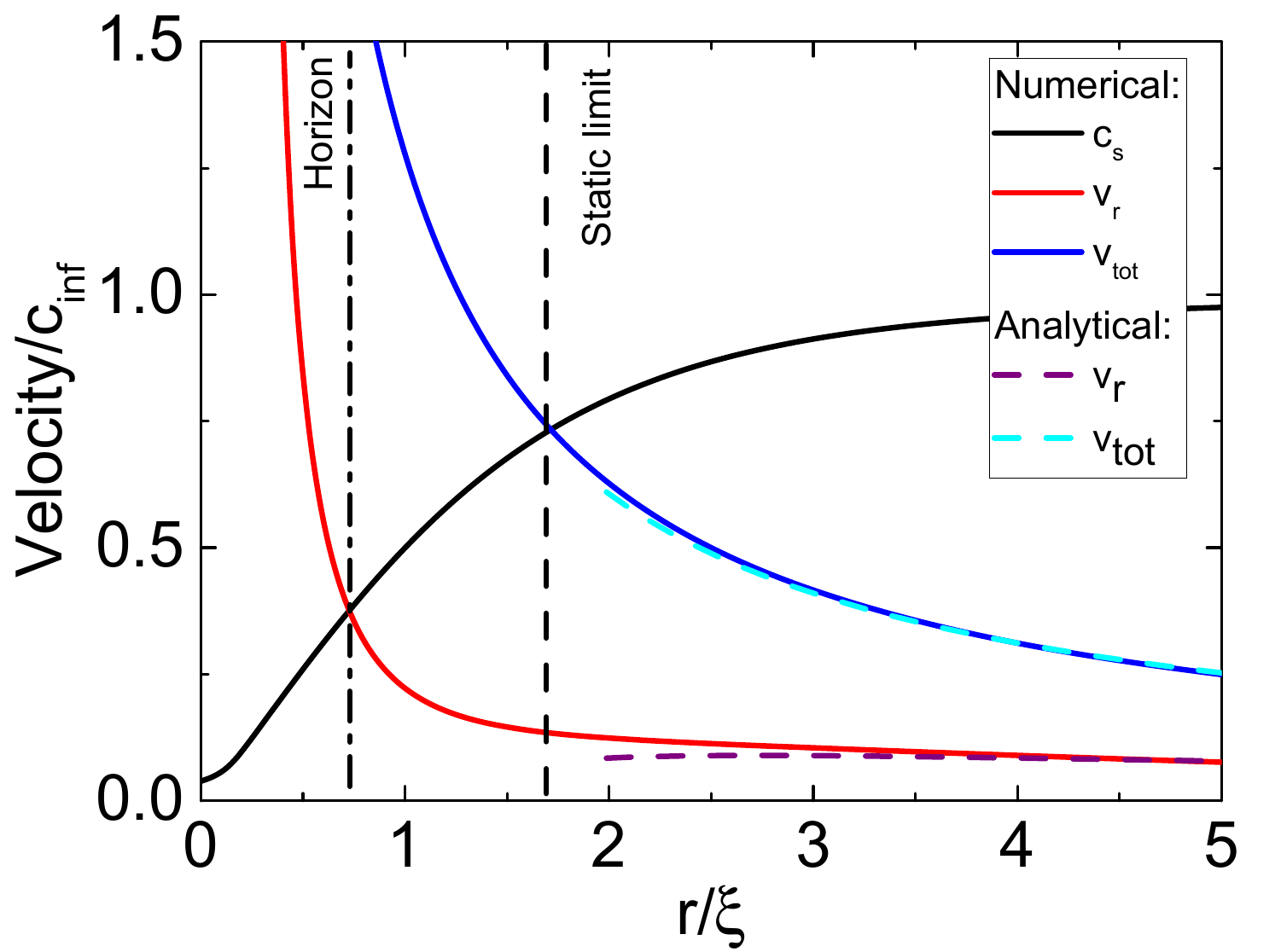}
    \caption{The speed of sound $c_s$, radial velocity $v_r$ and total velocity $v_{tot}$ as functions of $r$ (normalized by the sound velocity at infinity $c_{inf}$ and the healing length $\xi$). The positions of the horizon and the static limit (defined by $v_r=c_s$ and $v_{tot}=c_s$, respectively) are marked by vertical dashed lines.}
    \label{figProfiles}
\end{figure}

Figure~\ref{figProfiles} shows $v_{tot}$ and ${v_r}$. The numerical solution is plotted with solid lines: speed of sound $c_s$ (black), radial velocity $v_r$ (red), and total velocity $v_{tot}$ (blue). One clearly see the presence of an horizon and of a static limit.
Additional numerical simulations (not shown) indicate that the radius of the horizon, as well as the ratio of the horizon to the static limit, increase with $\Lambda$. In practice, $\Lambda$ can be controlled \cite{Solnyshkov2014}  by the temperature (independently from the polariton mass) or by the detuning (which also affects the polariton mass and the interaction constant).

We now turn to an analytical description of the system under study. Analytical solution for the whole vortex wavefunction is, unfortunately, unknown even in the conservative case. The description of the core in the non-conservative case turns out to be more complicated, so we focus on the opposite limit. The results are still useful, because the long distance interaction between vortices is provided by velocity profiles far from the core.

We therefore consider the limiting case $r\gg \xi$. At large distances, by applying series expansion to the equation for the radial part of the wavefunction \cite{Pitaevskii}, one can obtain an approximate solution $\psi  \approx  \sqrt n \left( {1 - {\xi ^2}/{r^2}} \right){e^{i\theta }}$. We then insert this solution into Eq.~\eqref{gpeloss}. The damping term gives two contributions, proportional to the Laplacian:
\begin{equation}
\Delta \psi  = \left( {\frac{1}{r}\frac{\partial }{{\partial r}}r\frac{\partial }{{\partial r}}\psi  + \frac{1}{{{r^2}}}\frac{{{\partial ^2}\psi }}{{\partial {\theta ^2}}}} \right)\sim \left( {\frac{{ - 4{\xi ^2}}}{{{r^4}}} - \frac{1}{{{r^2}}}} \right)\psi 
\end{equation}
The first contribution is due to the spatial variation of the density profile, while the second contribution is due to the centrifugal term, which is dominant for $r\gg \xi$. We obtain therefore:
\begin{equation}
    i\Lambda\frac{\hbar^2}{2m}\Delta\psi\approx -\frac{i\Lambda\hbar^2}{2mr^2}
\end{equation}
where $r$ is measured from the vortex center. We therefore find that the losses exhibit a spatial pattern determined by the position of the vortex: if the vortex moves, the loss pattern moves as well. This is like having a draining bathtub configuration with a mobile drain accompanying each vortex.

To find analytically the expression for the radial velocity of the flow for a single vortex with $k^2$-losses, we use the continuity equation: since the solution is stationary, all losses inside a certain circle with a radius $R$ are compensated by an inward flow through this circle. The total losses can be found as
\begin{equation}
    \Gamma\sim \int\limits_\xi^R \frac{1}{r^2}2\pi r\, dr \sim \log\frac{R}{\xi}
\end{equation}
Here, again, we do not consider the losses taking place inside the vortex core ($r<\xi$), where our approximation is not applicable. The compensation of these losses by the flow allows to find an asymptotic expression for the radial velocity, again valid for $r\gg\xi$:
\begin{equation}
    \Gamma=2\pi R v_r(R)
\end{equation}
which gives
\begin{equation}
    v_r\sim-\Lambda\frac{\log r/\xi}{r}
    \label{vr}
\end{equation}

As we will show below, in the next section, the strength of the attraction between vortices is determined by the radial velocity itself proportional to the damping coefficient $\Lambda$. If $\Lambda$ is sufficiently strong, it can overcome the well-known long-range repulsion~\cite{Pitaevskii} of same-sign vortices, allowing to observe their bound states.

Figure~\ref{figProfiles} compares the numerical solution of the equation~\eqref{gpeloss}  with the analytical estimates that we plot only in its validity range (for $r>2\xi$).  In this limit, the 
numerical solution is in a good agreement with the  behavior suggested by the analytics: it presents the same scaling $\log r/r$ for the radial velocity for $r\gg\xi$. The position of the static limit is predicted correctly. 
However, the analytical formula is not valid in the region $r\approx \xi$ where the event horizon is present and does not give any insight about its existence.

\subsection{Two bodies: zero angular momentum}

In the Newtonian picture, the material points cannot have their proper rotation. The only angular momentum possible is their angular momentum with respect to the center of mass. If it is zero (zero impact parameter), then the Earth falls directly on the Sun following a straight line. For a non-zero angular momentum, the bodies follow elliptical trajectories (in the bound configuration).

In relativity, each of the bodies can have its own angular momentum which cannot be neglected. The frame dragging associated with it leads to the Lense-Thirring effect~\cite{ciufolini2004confirmation}: the spacetime is "dragged" in the direction of rotation of a rotating body, and everything inside this spacetime is also "dragged" in the same direction. The straight infalling trajectories for zero "external" angular momentum thus become curved \cite{adler1966introduction} because of the "internal" momentum of the attracting body: the Earth does not follow a straight line anymore, but still falls on the Sun for zero impact parameter.

\begin{figure}
    \centering
    \includegraphics[width=0.8\linewidth]{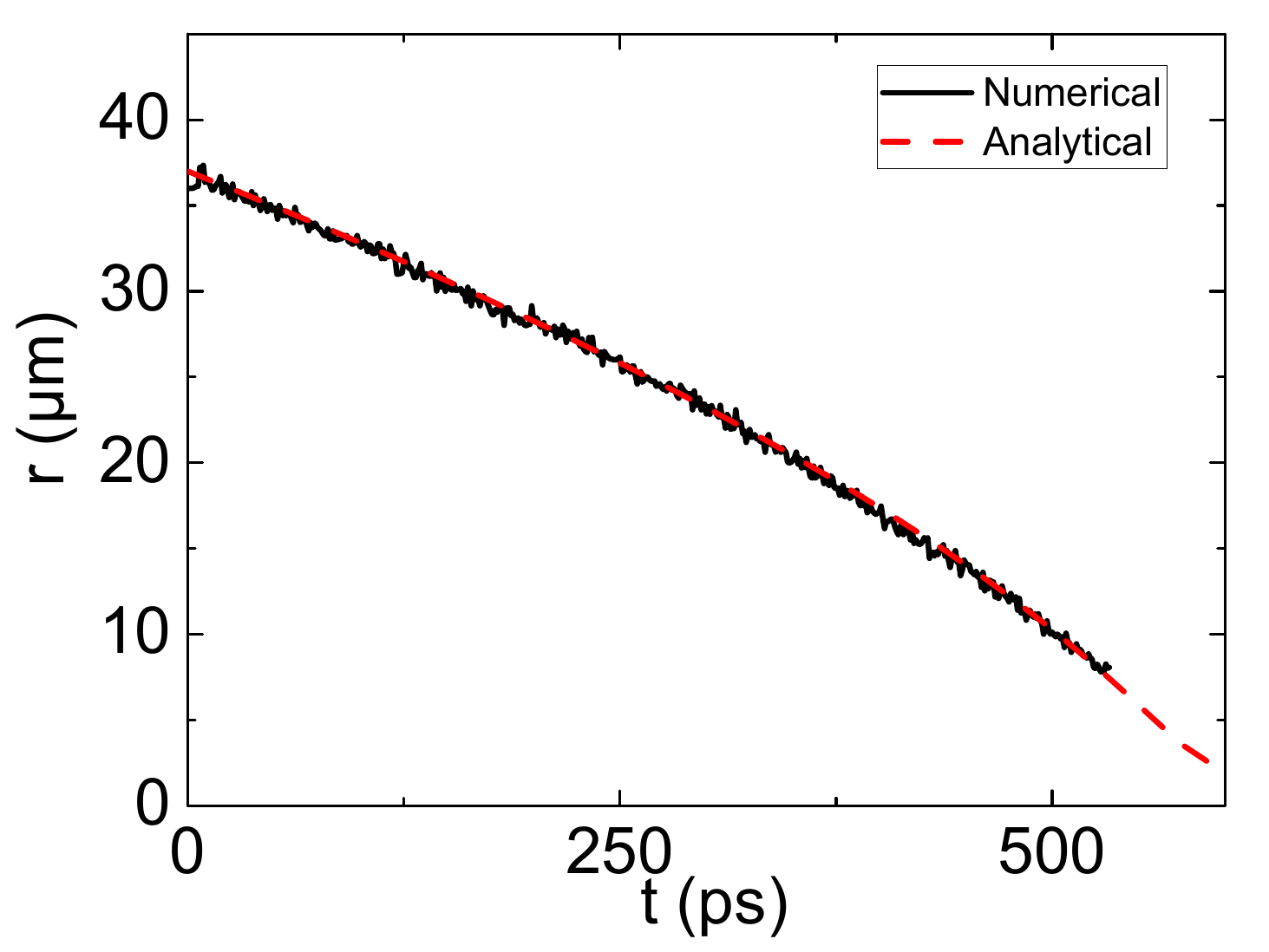}
    \caption{Zero angular momentum merger: time evolution of the distance between the analogue black holes. Numerical simulation (black line) and analytical solution~\eqref{zamsol} (red dashed line). }
    \label{fig1}
\end{figure}

In the most simple case, each of the vortices is moving with the flow surrounding its core, which is defined by the velocity pattern created by the other vortex. In this case, the "external" angular momentum is zero, and all the rotation that can be observed is due to the frame-dragging ("internal" angular momentum of each vortex).
In this case, the radial velocity given by Eq.~\eqref{vr} completely determines the radial motion of each of them via the following differential equation:
\begin{equation}
    \frac{dr}{dt}=-a\frac{\log r/\xi}{r}
\end{equation}
where $a$ is an adjustable parameter (a proportionality constant, which includes $\Lambda$ from Eq.~\eqref{vr}).
The solution of this differential equation can be explicitly written for $t(r)$: 
\begin{equation}
    t(r)=t_0-\xi^2 a^{-1}\mathrm{Ei}\left(2\log \frac{r}{\xi}\right)
    \label{zamsol}
\end{equation}
with a special function (exponential integral) defined as
\begin{equation}
    \mathrm{Ei}(z)=-\int\limits_{-z}^\infty\frac{e^{-t}}{t}dt
\end{equation}
This analytical solution is compared with numerical simulations for a pair of vortices in Fig.~\ref{fig1}. The black line shows the distance between the centers of the two vortices from numerical simulations, and the red dashed line shows a fit based on the analytical model~\eqref{zamsol}.

This configuration corresponds to a collision of two black holes with the same proper rotation but with zero external angular momentum (zero impact parameter). No emission of gravitational waves is required for the vortices to attract at large distances because there is no angular momentum and associated rotational energy to dissipate. However, mutual rotation of the black holes can still be observed due to frame dragging, that is, the rotation of the spacetime itself in the direction of the rotation of each black hole. It is indeed observed in numerical simulations.

\subsection{Non-zero angular momentum: Kepler's law}

We now turn to the case of non-zero external angular momentum. In this case, in the framework of Newtonian celestial mechanics, if the emission of gravitational waves is neglected, each of the two bodies is expected to follow an elliptic orbit forever, with the period of rotation given by the 3rd Kepler's law. To be able to describe the effect of the gravitational waves in our analogue system, we need to find an equivalent of this law first.

\begin{figure}
    \centering
    \includegraphics[width=0.8\linewidth]{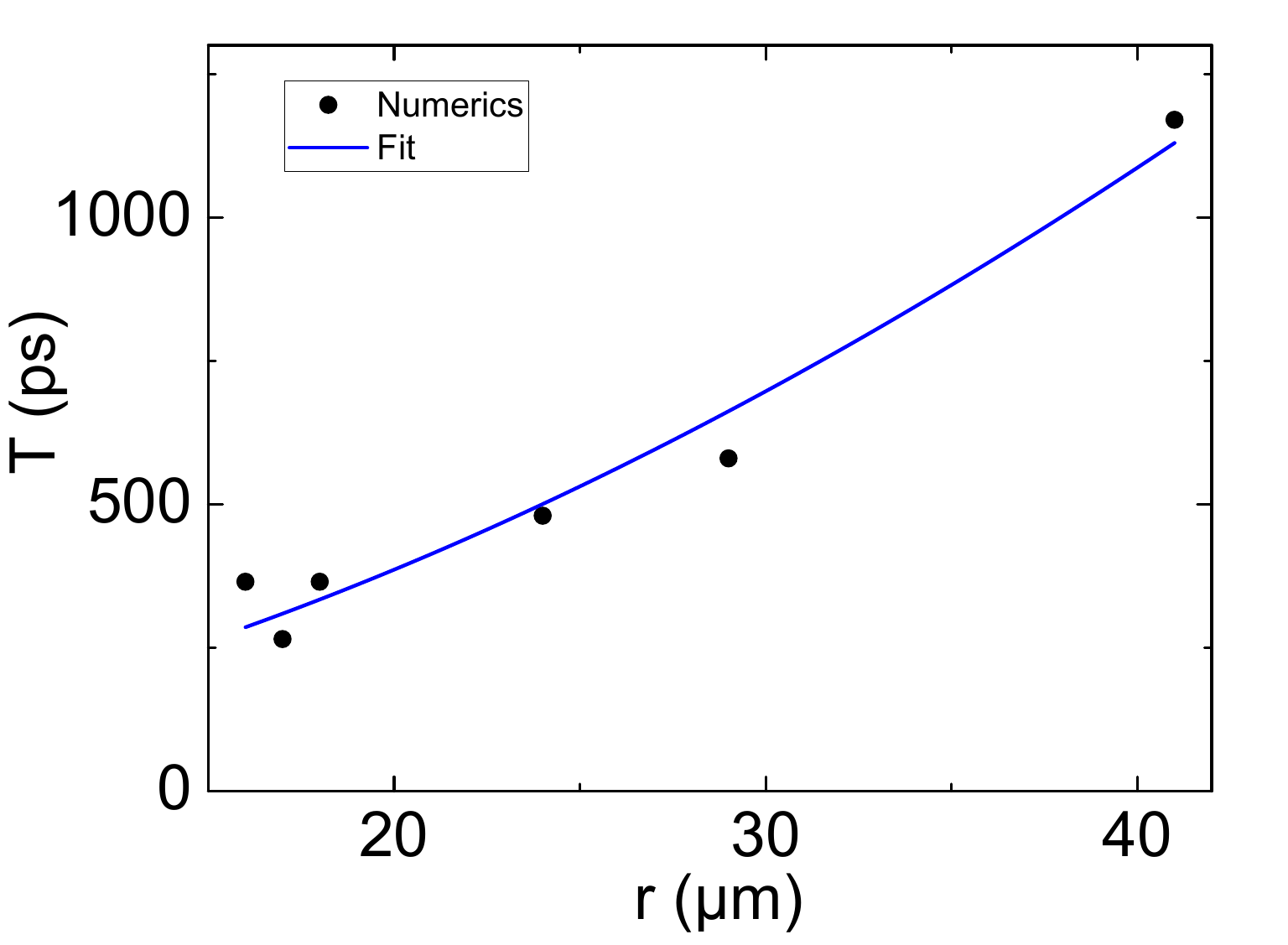}
    \caption{Rotation period as a function of distance between the analogue black holes: black dots - numerical simulations, blue solid line - analytical solution~\eqref{Keplaw}. }
    \label{fig2}
\end{figure}

For $r\gg \xi$, $\log r/\xi$ is a slowly varying function with respect to $r$, and therefore we can assume that $\alpha=\log r/\xi$ is approximately constant for a single rotation period. We can thus consider a simplified differential equation for the radial coordinate
\begin{equation}
    \frac{dr}{dt}=-\frac{\alpha}{r}
\end{equation}
whose solution is $r(t)=\sqrt{-2at}$. This allows us to find the radial acceleration
\begin{equation}
\frac{d^2r}{dt^2}\sim \frac{\alpha^2}{r^3}    
\end{equation}
which allows us to conclude that the attractive force (which is responsible for this acceleration) behaves as 
\begin{equation}
    F(r)\sim \frac{\alpha^2}{r^3}\sim \frac{\log r/\xi}{r^3}
\end{equation}
and the corresponding potential energy is 
\begin{equation}
    U(r)\sim -\alpha^2/r^2
    \label{upot}
\end{equation}
This potential energy is decreasing faster than the Newton's law of gravity in 3D.
Then, using the fact that the centripetal acceleration for stationary circular motion (considering that the orbit is approximately circular) should be $d^2r/dt^2=v_\theta^2/r$, we can find the azimuthal velocity
\begin{equation}
    v_\theta\sim \frac{\alpha}{r}
    \label{vaz}
\end{equation}
which gives the rotation period
\begin{equation}
    T\sim \frac{r^2}{\alpha}=\frac{r^2}{\log r/\xi}
    \label{Keplaw}
\end{equation}
This is not a simple power law but has two asymptotics.
For $r\gg\xi$, $T\sim r^2$, but for $r\sim \xi$ the logarithm cannot be neglected and $T\sim r$, so the power law changes between $r^1$ and $r^2$. This differs from the 3rd Kepler's law: $T\sim r^{3/2}$, but all three power law exponents are of the same order. 

These analytical results are compared with the results of numerical simulations for a pair of vortices in Fig.~\ref{fig2}: the blue solid line shows the modified Kepler's law, while the black dots are the results of numerical simulations. The numerical results are quite well described by the approximate analytical solution.

\subsection{Gravitational waves}

\begin{figure}
    \centering
    \includegraphics[width=0.8\linewidth]{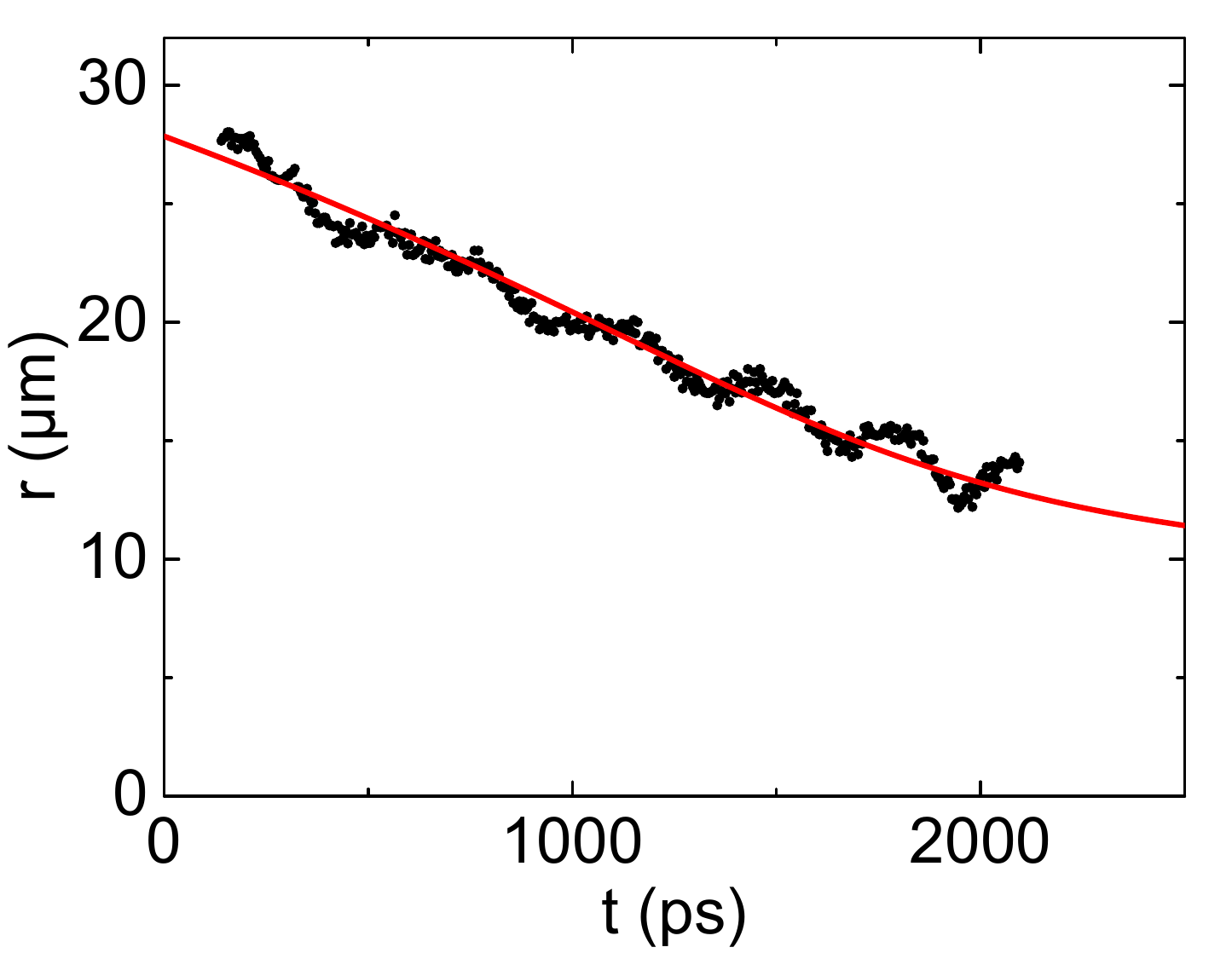}
    \caption{Non-zero angular momentum merger: time evolution of the distance between the analogue black holes. Black dots - numerical simulation, red solid line - analytical solution~\eqref{nzsol}.}
    \label{fig3}
\end{figure}

We now study the energy loss suffered by an accelerated vortex via the emission of waves. The nature of these waves is dual: the density waves in the condensate can at the same time be considered as gravitational waves (because they modify the metric of the condensate, which is density-dependent), but also as electromagnetic waves (because the vortices behave as charges \cite{Popov1972} interacting via an electromagnetic-like field). Since the dipole mechanism of wave emission is suppressed by the momentum conservation law, the strongest emission corresponds to the quadrupole moment. The expression for the intensity of the quadrupole emission is universal, and does not depend on the nature of the waves (electromagnetic vector waves or gravitational tensor waves) \cite{Landau2}  :
\begin{equation}
    \frac{dE}{dt}\sim -\dddot{Q}^2
\end{equation}
where the quadrupole moment of a pair of bodies on a circular orbit is
\begin{equation}
    Q_{xx}=mr^2\left(3\cos^2\theta -1\right)
\end{equation}
(other components are similar), and its 3rd-order time derivative therefore behaves as
\begin{equation}
    \dddot{Q}_{xx}=24mr^2\omega^3\cos\omega t\sin\omega t\sim r^2\omega^3
\end{equation}
where $\omega$ is the rotation frequency.

Using the centripetal acceleration determined by the azimuthal velocity \eqref{vaz}, we obtain
\begin{equation}
    \frac{dE}{dt}\sim -r^4\omega^6\sim -\frac{\alpha^6}{r^8}
    \label{eloss}
\end{equation}
Combining the energy loss~\eqref{eloss} with the potential energy~\eqref{upot}, we find the following differential equation
\begin{equation}
    \frac{d}{dt}\left(-\frac{\alpha^2}{r^2}\right)\sim -\frac{\alpha^6}{r^8}
\end{equation}
where we neglect the time derivative of $\alpha$ because of the slow variation of $\log r/\xi$, as before. This allows us to simplify and rewrite the equation as:
\begin{equation}
    \frac{r^5}{\log^5 r/\xi}dr\sim dt
\end{equation}
The solution for the trajectory can finally be written as an inverse function $t(r)$:
\begin{equation}
t = {t_0} - A\left( {54{\xi ^6}{\mathop{\rm Ei}\nolimits} \left( {6\log \frac{r}{\xi }} \right) - {r^6}\left( {\frac{1}{{4{{\log }^4}\frac{r}{\xi }}} + \frac{1}{{2{{\log }^3}\frac{r}{\xi }}} + \frac{3}{{2{{\log }^2}\frac{r}{\xi }}} + \frac{9}{{\log \frac{r}{\xi }}}} \right)} \right)
    \label{nzsol}
\end{equation}
For large distances, the variation of the $\log r$ terms can be neglected, and the solution scales as $r(t)\sim (t_0-t)^{1/6}$: it is a power law behavior, as for the inspiral of the astrophysical black holes \cite{LIGO2016}, but with a different power ($1/6$ vs $1/4$). For shorter distances, the logarithms and the exponential integral functions start to play an important role, and the resulting trend becomes qualitatively different: at large $t$, $r-\xi\sim 1/t$. This is a consequence of the analytical approximation, which is not valid for $r\sim\xi$.

This solution is compared with the results of a numerical experiment in Fig.~\ref{fig3}. The analytical solution is plotted as a solid line, while the experimental points are shown as black dots, with a good agreement between the two. The average decrease of the distance is much slower than in the case with zero external angular momentum shown in Fig.~\ref{fig1}. It also seems that the vortices cannot approach each other closer than a certain distance, due to the modification of the spatial profiles of their cores, as compared to the case of a single vortex seen in Fig.~\ref{figProfiles}. This leads to repulsion at short distances, which overcomes the attraction. The short range interaction and the means to overcome the repulsion will be studied in future works.

\begin{figure}
    \centering
    \includegraphics[width=0.95\linewidth]{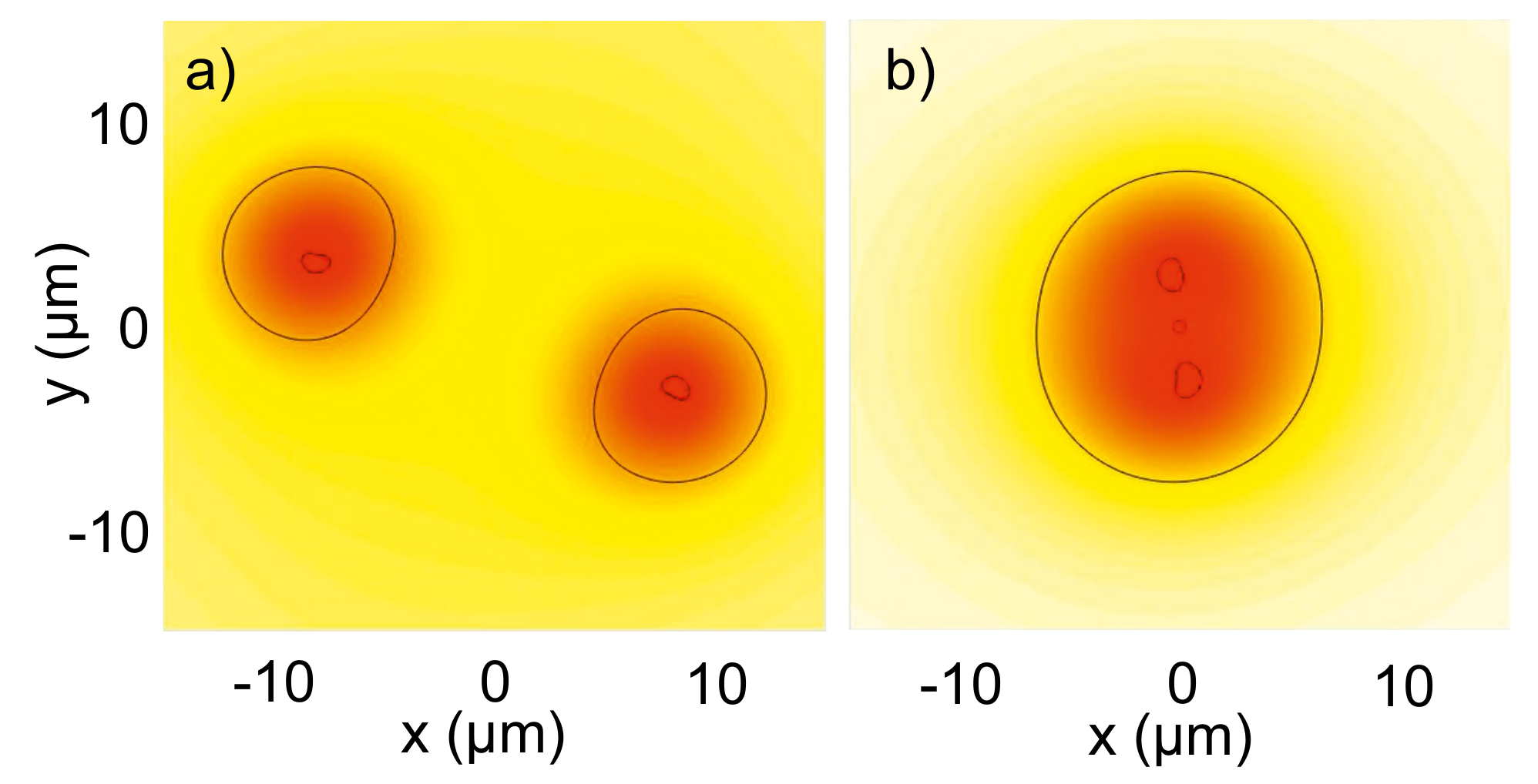}
    \caption{Two snapshots of the analogue black holes at the early (a) and late (b) stages of the inspiral. False color shows the polariton density $|\psi|^2$, black lines mark the position of the static limit and the horizon.}
    \label{figMerg}
\end{figure}

Finally, in Fig.~\ref{figMerg} we show the snapshots of the polariton density distributions at two different moments of time corresponding to early (panel a) and late (panel b) stages of the inspiral. The horizon and the static limit are shown with black lines.  So far, our simulations did not allow us to reach the merger of the horizons, but we hope that it could be possible under different conditions (stronger losses or a different $k$-dependence of these losses). On the contrary, a merger of the static limits does occur already in the simulation shown in Fig.~\ref{figMerg}.

We note that from symmetry considerations, the velocity in the center of the system (between the two vortices) is necessarily zero, and thus this point cannot be "inside" an analogue Kerr black hole, whose interior is defined by the condition $v_r>c$. On the other hand, real black holes, as for example in the case of Kerr-Newman metric, can have a complicated internal structure with several horizons \cite{newman1965metric,mcnamara1978instability,brady1999internal}, and similar effects occur in analogue black holes \cite{Steinhauer2014,kolobov2021observation} (so far only in 1D). We can therefore conclude that if a merger is achieved, the resulting configuration could have two concentric horizons.

\section{Conclusions}

To conclude, we have demonstrated that the presence of a $k^2$-loss term in quantum fluid creates a convergent current for quantum vortices. This type of term is naturally present in exciton-polariton quantum fluid,  either freely decaying, if the exciton loss exceeds the photon loss, or non-resonantly pumped, when an equilibrium Bose-Einstein condensate forms. The metric associated with quantum vortex is similar to that of a Kerr black hole, showing both a horizon and a static limit. We show that the convergent "draining bathtub" flow attached to each vortex provokes an attraction between them and leads to the emergence of an analogue of attractive gravitational interaction for vortices. For a pair of vortices, we demonstrate how the emission of analogue gravitational waves leads to the inspiral of these analogue black holes, with increasing rotation frequency.

\bibliographystyle{crunsrt}

\bibliography{biblio}
\end{document}